\documentclass[amsmath,aps,showpacs,a4paper,10pt]{revtex4}

 \usepackage{epsf}
 \usepackage{graphicx}    

 \usepackage{amssymb,amsmath,amscd}     

\begin{document}

 \newcommand{\be}[1]{\begin{equation}\label{#1}}
 \newcommand{\ee}{\end{equation}}
 \newcommand{\bea}{\begin{eqnarray}}
 \newcommand{\eea}{\end{eqnarray}}
 \def\disp{\displaystyle}

 \def\gsim{ \lower .75ex \hbox{$\sim$} \llap{\raise .27ex \hbox{$>$}} }
 \def\lsim{ \lower .75ex \hbox{$\sim$} \llap{\raise .27ex \hbox{$<$}} }

 \begin{titlepage}

 \begin{flushright}
 arXiv:1601.01625
 \end{flushright}

 \title{\Large \bf Stability of Differentially Rotating Disks
 in $f(T)$ Theory}

 \author{Shoulong~Li\,}
 \thanks{\,email address:\ sllee$_{-}\hspace{-0.275mm}$phys@bit.edu.cn}
 \affiliation{School of Physics,
 Beijing Institute of Technology, Beijing 100081, China}

 \author{Hao~Wei\,}
 \thanks{\,Corresponding author}
 \email[\,email address:\ ]{haowei@bit.edu.cn}
 \affiliation{School of Physics,
 Beijing Institute of Technology, Beijing 100081, China}

 \begin{abstract}\vspace{1cm}
 \centerline{\bf ABSTRACT}\vspace{2mm}
 To explain the accelerated expansion of our universe, many
 dark energy models and modified gravity theories have been
 proposed so far. It is argued in the literature that they
 are difficult to be distinguished on the cosmological scales.
 Therefore, it is well motivated to consider the relevant
 astrophysical phenomena on (or below) the galactic scales.
 In this work, we study the stability of self-gravitating
 differentially rotating galactic disks in $f(T)$ theory, and
 obtain the local stability criteria in $f(T)$ theory, which
 are valid for all $f(T)$ theories satisfying $f(T=0)=0$ and
 $f_T (T=0)\not=0$, if the adiabatic approximation and the weak
 field limit are considered. The information of the function
 $f(T)$ is mainly encoded in the parameter
 $\alpha\equiv 1/f_T(T=0)$. We find that the local stability
 criteria in $f(T)$ theory are quite different from the ones
 in Newtonian gravity, general relativity, and other modified
 gravity theories such as $f(R)$ theory. We consider that this
 might be a possible hint to distinguish $f(T)$ theory from
 general relativity and other modified gravity theories on
 (or below) the galactic scales.
 \end{abstract}

 \pacs{04.50.Kd, 98.80.-k, 04.40.-b, 98.62.-g}

 \maketitle

 \end{titlepage}

 \renewcommand{\baselinestretch}{1.0}


\section{Introduction}\label{sec1}

The discovery of accelerated expansion of our universe from
 the observations of Type Ia supernovae in 1998~\cite{Riess:1998cb}
 has been one of the most amazing achievements in modern
 cosmology. The simplest way to explain this fantastic phenomenon is
 introducing a cosmological constant~\cite{Padmanabhan:2002ji,cc1}.
 However, it is plagued with the fine-tuning and cosmic coincidence
 problems (see e.g.~\cite{Padmanabhan:2002ji}). So, many
 dynamical dark energy (DE) models have been proposed, for instance,
 quintessence~\cite{qt1}, phantom~\cite{ph1} and so on. On the other
 hand, modifying general relativity (GR)~\cite{mg1} on the
 cosmological scales has also been extensively considered to explain
 the accelerated expansion of our universe without introducing
 DE. In fact, there exist many modified gravity theories in the
 literature, such as $f(R)$ theory~\cite{fr1},
 $f(T)$ theory~\cite{Bengochea1,Linder,Ferraro:2006jd}, massive
 gravity~\cite{mg2}. In particular, the well-known $f(R)$
 theory~\cite{fr1} is constructed by replacing the Einstein-Hilbert
 Lagrangian (namely Ricci scalar $R$) with an arbitrary
 function $f(R)$. Similarly, the so-called $f(T)$
 theory~\cite{Bengochea1,Linder,Ferraro:2006jd}
 is constructed by replacing the ordinary Lagrangian (namely
 torsion scalar $T$) in the teleparallel equivalent of general
 relativity (TEGR) with an arbitrary function $f(T)$. In fact,
 there are other modified gravity theories in the literature,
 such as scalar-tensor theory, braneworld model, Galileon
 gravity, Gauss-Bonnet gravity, etc. We refer to
 e.g.~\cite{mg1,fr1,Bengochea1,Linder,mg2,Cai:2015emx,
 Ferraro:2006jd} for more details about modified gravity.

Since we have a flood of models of DE and modified gravity in
 the literature, some authors focus on how to differentiate one
 model from others. For example, in order to differentiate
 dynamical DE from cosmological constant, the expansion history
 of the universe is considered. Caldwell and Linder proposed a
 so-called $w-w'$ analysis in~\cite{Caldwell1}, and then it was
 extended in e.g.~\cite{Scherrer1}. Another method is the
 statefinder diagnostic proposed by Starobinsky {\it et
 al.}~\cite{Sahni1}. We refer to e.g.~\cite{Wei1}
 and references therein for some relevant works on $w-w'$
 analysis and statefinder diagnostic. However, as is well known
 (see e.g.~\cite{Sahni:2006pa}), one can always build models
 sharing a same cosmic expansion history, and hence these
 models cannot be distinguished by using the expansion history only.
 Later, it is realized that if the cosmological models share a same
 cosmic expansion history, they might have different growth
 histories characterized by the growth function
 $\hat{\delta}(\hat{z})\equiv\delta\rho_m/\rho_m$ (namely the
 matter density contrast as a function of redshift $\hat{z}$).
 Therefore, the cosmological models might be distinguished from
 each other by combining the observations of both the expansion and
 growth histories (see e.g.~\cite{Wei2}). However, this approach has
 been challenged. For instance, if non-trivial dark energy
 clustering (e.g. non-vanishing anisotropic stress) is allowed,
 it is found in e.g.~\cite{Kunz:2006ca,Bertschinger:2008zb}
 that DE models still cannot be distinguished from modified
 gravity theories even by using the observations of both the
 expansion and growth histories. On the other hand, without
 invoking non-trivial dark energy clustering, it is found in
 e.g.~\cite{Wei3} that the interacting DE models also cannot
 be distinguished from modified gravity theories. In fact,
 the work in~\cite{Wei3} was further generalized
 in~\cite{Wei4}. It is found that the interacting DE models,
 modified gravity theories and warm dark matter models are
 indistinguishable~\cite{Wei4}. Therefore, the complementary
 probes beyond the ones of cosmic expansion history and
 growth history are required to distinguish
 various cosmological models.

Notice that in all the arguments of~\cite{Wei2, Kunz:2006ca,
 Bertschinger:2008zb,Wei3,Wei4}, it is assumed that the matter
 density contrast $\hat{\delta}(\hat{z})$ is linear. If this
 key assumption is invalid, the hope to distinguish various
 cosmological models still exists. This draws our attention to
 the observations on the smaller scales. For instance, the
 galactic scales might be suitable, since the matter density
 contrast $\hat{\delta}(\hat{z})$ becomes non-linear on (or
 below) these scales. So, it is well motivated to consider the
 relevant astrophysical phenomena on (or below) the galactic scales.

In the present work, we are interested to consider the
 stability of equilibria of galactic disks. We hope this can
 bring some possible hints to discriminate various modified
 gravity models. The study of stability of stellar equilibria
 is one of the most important tasks in galactic
 dynamics~\cite{Binney1}, since it is used in many aspects such
 as the star formation~\cite{Quirk1972}. For many equilibria of
 stellar systems, a slight perturbation will cause them evolve
 violently away from their initial states. In other words, many
 equilibria of stellar systems are unstable. The instabilities
 considered in this work are caused by cooperative effects, in
 which a density perturbation gives rise to extra gravitational
 forces deflecting the stellar orbits in such a way that
 the original density perturbation is enhanced. In order to
 relate to observation, the stellar systems considered in this
 work is differentially rotating disks, because realistic
 galactic disks are not static and do not rotate uniformly. Besides,
 the phenomena of instability in disks are strongly influenced
 by differential rotation. The stability analysis for differentially
 rotating systems are more difficult than static spherical
 systems and uniformly rotating systems. Fortunately, Lin and
 Shu~\cite{LS} recognized that the structure of spiral arms in
 a stellar disk could be regarded as a density wave, a wavelike
 oscillation that propagates through the disk in much the same
 way that waves propagates through violin strings. They also
 realized that the techniques of wave mechanics, which has been
 developed into the so-called density wave theory, could be
 applied to deduce the properties of density
 waves in differentially rotating stellar disks.

To our knowledge, the stability analysis was firstly investigated by
 Safronov~\cite{Safronov1} and Toomre~\cite{Toomre1}, respectively.
 By studying the behavior of the density waves in
 self-gravitating disks in Newtonian gravity, Toomre~\cite{Toomre1}
 derived the so-called dispersion relation which relates the
 wavenumber to its frequency, and then he got the local
 stability criterion (known as Toomre's stability criterion in
 the literature) which can be used to approximately determine
 whether the system is locally stable to a small axisymmetric
 perturbation by applying the dispersion relation. It is worth
 noting that the gravitational potential plays an important
 role in deriving the dispersion relation and local stability
 criterion for a given system. When we study the stability in
 GR, the derived gravitational potential  in the weak field
 limit reduces to the one in Newtonian gravity. However, in the
 framework of modified gravity theory rather than GR, it is natural
 to anticipate a different gravitational potential, and then
 a different dispersion relation as well as a different local
 stability criterion for a given self-gravitating system.
 Therefore, the stability of self-gravitating disk might be used to
 reflect the difference between various gravity theories.

In the literature, the stability of differentially rotating
 disks were considered in some modified gravity theories,
 e.g. Modified Newtonian Dynamics (MOND)~\cite{Milgrom},
 Moffat's Modified Gravity (MOG)~\cite{Roshan1,Roshan:2014mqa},
 $f(R)$ theory~\cite{Roshan2}, and so on. In the present work,
 we try to consider the stability of differentially rotating
 disks in another modified gravity theory, namely $f(T)$
 theory~\cite{Bengochea1,Linder,Ferraro:2006jd}. As one of the
 important modified gravity theories, $f(T)$ theory has some
 interesting features. For instance, unlike the metric $f(R)$
 theory whose equations of motion are 4th order, the equations
 of motion in $f(T)$ theory are 2nd order. This virtue makes
 $f(T)$ theory fairly attractive. On the other hand, it is
 worth noting that $f(T)$ theory can provide a mechanism for
 realizing bouncing cosmology~\cite{Cai:2011tc}, thereby
 avoiding the Big Bang singularity. The seminal
 work of~\cite{Ferraro:2006jd} also obtained a cosmological
 picture without the initial singularity. This is another one
 of the advantages of $f(T)$ theory. In fact, there are
 many relevant works on $f(T)$ theories in the
 literature (e.g.~\cite{Wu:2010mn,Wei:2011mq,Bamba:2010wb,
 Wei:2011aa,Wu:2010xk,Chen:2010va,Li:2010cg,Dent:2011zz,Cai:2011tc,
 Ferraro:2012wp,Tamanini:2012hg}). Actually $f(T)$ theory
 becomes one of the active fields in cosmology, and it received
 a huge attention from the very beginning, as is shown by the
 references in~\cite{Cai:2015emx}.

The rest of this paper is organized as follow. In
 Sec.~\ref{sec2}, we briefly review the key points of $f(T)$
 theory, and then derive the Poisson's equation in the weak field
 limit. In Sec.~\ref{sec3}, we consider the behavior of density
 waves in differentially rotating disks in $f(T)$ theory by
 using the density wave theory. We obtain the perturbed
 gravitational potential by considering a small perturbation of
 the equilibria, and derive the dispersion relations for both
 gaseous disks and stellar disks. In Sec.~\ref{sec4}, we use
 the dispersion relations to get the corresponding Toomre's
 local stability criteria for an axisymmetric perturbation in
 both gaseous disks and stellar disks. Finally, some brief
 concluding remarks are given in Sec.~\ref{sec5}.


\section{$f(T)$ theory}\label{sec2}


\subsection{Key points of $f(T)$ theory}\label{sec2A}

Here, we briefly review $f(T)$ theory following
 e.g.~\cite{Bengochea1,Linder,Ferraro:2006jd}. In fact, it is
 a generalization of TEGR. Teleparallelism uses
 the Weitzenb\"ock connection~\cite{Weitzen1923} (which has
 torsion but it is curvatureless) to describe spacetime, while
 GR uses the Levi-Civita affine connection (which has curvature
 but it is torsionless). In this sense, teleparallelism
 is a sector of Einstein-Cartan theories~\cite{Cartan1922,
 Hehl:1976kj}, which describe gravity by means of a connection
 having both torsion and curvature. Teleparallel gravity uses a
 vierbein field $e_a = {e_a}^\mu \partial_\mu$ as dynamical
 quantity, where Greek indices $\mu$, $\nu$, $\ldots$, run over
 $0,\,1,\,2,\,3$ on the manifold, and Latin indices $a$, $b$,
 $\ldots$, run over $0,\,1,\,2,\,3$ on the tangent space of the
 manifold. We use the Einstein notation throughout this work.
 The vierbein field ${e^a}_\mu$ relates to the spacetime metric
 $g_{\mu\nu}$ through~\cite{Bengochea1,Linder,Ferraro:2006jd}
 \be{eq5a}
 g_{\mu\nu}=\eta_{ab}\, {e^a}_\mu \, {e^b}_\nu \,,
 \ee
 where $\eta_{ab}= {\rm diag}\,(+1,\,-1,\,-1,\,-1)$ is the
 Minkowski metric. The Weitzenb\"ock connection
 ${\Gamma^\lambda}_{\mu\nu}$ and the
 torsion tensor ${T^\rho}_{\mu\nu}$ are defined
 by~\cite{Bengochea1,Linder,Ferraro:2006jd}
 \bea
 &{\Gamma^\lambda}_{\mu\nu} =
 {e_a}^{\lambda}\,\partial_\mu {e^a}_\nu \,,\nonumber\\[1mm]
 &{T^\rho}_{\mu\nu} = {\Gamma^\rho}_{\nu\mu}-{\Gamma^\rho}_{\mu\nu}
 = -{e_a}^\rho \left(\partial_\mu\,{e^a}_\nu
 -\partial_\nu\,{e^a}_\mu \right) \,.\label{eq3}
 \eea
 Unlike the Levi-Civita connection, the indices $\mu$ and $\nu$
 in Weitzenb\"ock connection are not symmetric, so the torsion
 tensor is not vanishing. The torsion scalar is
 given by~\cite{Bengochea1,Linder,Ferraro:2006jd}
 \be{eq2}
 T= {T^\rho}_{\mu\nu}\,{S_\rho}^{\mu\nu} \,,
 \ee
 where
 \be{eq4}
 {S_\rho}^{\mu\nu} = \frac{1}{2} \left( {K^{\mu\nu}}_\rho +
 \delta^\mu_\rho\, {T^{\lambda\nu}}_\lambda - \delta^\nu_\rho\,
 {T^{\lambda\mu}}_\lambda \right)\,,
 \ee
 and ${K^{\mu\nu}}_\rho$ is contorsion tensor,
 \be{eq5}
 {K^{\mu\nu}}_\rho = -\frac{1}{2} \left( {T^{\mu\nu}}_\rho
 - {T^{\nu\mu}}_\rho-{T_\rho}^{\mu\nu}\right)\,.
 \ee
 The action of $f(T)$ theory is
 given by~\cite{Bengochea1,Linder,Ferraro:2006jd}
 \be{eq1}
 {\cal S}=\int d^4 x \left[\frac{\,|e|\,f(T)}
 {16\pi G}+{\cal L}_M\right]\,,
 \ee
 where $G$ is the gravitational constant, $|e|$ is the
 determinant of vierbein ${{e^a}_\mu}$, $f(T)$ is a function
 of torsion scalar $T$, and ${\cal L}_M$ is the Lagrangian of
 matter. Note that it reduces to TEGR if $f(T)=T$. We obtain
 the equations of motion for the vierbein
 from the variation of the action~(\ref{eq1}) with respect
 to ${e^a}_\nu$~\cite{Bengochea1,Linder,Ferraro:2006jd},
\be{eq6}
|e|^{-1}\,\partial_\mu\,\left( |e|\,{S_a}^{\mu\nu} \right)\, f_T(T)
 +\frac{1}{4}\,{e_a}^\nu\, f(T) + {S_a}^{\mu\nu}\,\partial_\mu
 \,T\,f_{TT}(T) + {T^\rho}_{\mu a}\,{S_\rho}^{\nu\mu}\,f_T(T) =
 4\pi G {e_a}^\mu \,{{\cal T}_\mu}^\nu \,,
\ee
where ${S_a}^{\mu\nu} = {S_\rho}^{\mu\nu}\,{e_a}^\rho$,
 ${T^\rho}_{\mu a} = {T^\rho}_{\mu\nu}\, {e_a}^\nu$, a subscript $T$
 denotes a derivative with respect to $T$, and ${\cal{T}_\mu}^\nu =
 -|e|^{-1} {e^a}_\mu \left({\delta {\cal L}_M}/{\delta
 {e^a}_\nu}\right)$ is the energy-momentum tensor of matter.
 In this work, we consider the perfect fluid, and
 its energy-momentum tensor is given by
 ${\cal T}^{\mu\nu}= -p g^{\mu\nu}+\left(p+\rho\right) u^\mu u^\nu$,
 where $\rho$, $p$ and $u$ are the energy density,
 pressure and velocity four-vector, respectively, with $u^0 = 1$ and
 $u^i = 0$. In the literature (e.g.~\cite{Bengochea1,Linder,
 Cai:2015emx}), some typical forms of $f(T)$ have
 been extensively considered, for instance,
 \be{eq16}
 f(T)= T + c_1\left(-T\right)^n\,,
 \ee
 and
 \be{eq17}
 f(T)= T + c_2\left(1-e^{-c_3\sqrt{-T}\,}\right)\,,
 \ee
 where $c_1$, $c_2$, $c_3$ and $n$ are constants. As is shown
 in the literature, these two specific $f(T)$ models are
 interesting and viable. For example, the constraints on these
 two typical $f(T)$ models were
 considered in e.g.~\cite{Wu:2010mn,Wei:2011mq}; the
 equation of state in these two $f(T)$ theories was studied in
 e.g.~\cite{Bamba:2010wb}; the Noether and Hojman symmetries
 were considered in e.g.~\cite{Wei:2011aa}; the dynamical
 behavior was studied in e.g.~\cite{Wu:2010xk}, and so on.
 In the literature, there are many relevant works on these
 two typical $f(T)$ models, and it is hard to mention them in
 details. We refer to e.g.~\cite{Cai:2015emx} (and the
 references therein) for a comprehensive review on $f(T)$
 theory. We stress that our results are not only valid for the
 specific forms of $f(T)$ given in Eqs.~(\ref{eq16}) and
 (\ref{eq17}). In fact, the results obtained in the present
 work do not rely on the specific forms of $f(T)$, and they
 are valid for all $f(T)$ theories satisfying $f(T=0)=0$ and
 $f_T (T=0)\not=0$, if the adiabatic approximation and the
 weak field limit are considered (see below).


\subsection{Weak field limit of $f(T)$ theory}\label{sec2B}

Here, we consider the weak field limit of $f(T)$ theory.
 Following e.g.~\cite{Binney1,Roshan2}, we restrict ourselves to the
 adiabatic approximation, in which the evolution of the universe is
 very slow in comparison with local dynamics. It means that we
 can choose the Minkowski metric $\eta_{ab}$ instead of
 Friedmann-Robertson-Walker~(FRW) metric as the background
 metric~\cite{Roshan2}. In fact, the quantitative studies
 (e.g.~\cite{bdsyspast,bdsysrecent,Nesseris:2004uj}) showed
 that the physics of gravitationally bound systems (such as
 galaxies, clusters, or planetary systems) which are small
 compared to the radius of curvature of the cosmological background
 is essentially unaffected by the expansion of the universe.
 Here, we briefly mention the key points of these quantitative
 studies (e.g.~\cite{bdsyspast,bdsysrecent,Nesseris:2004uj}).
 If the gravitationally bound system is massive enough, the
 local spacetime might be affected, and hence it should take an
 interpolating metric. By calculating the effective potential
 minimum, one can find the numerical evolution of the radius of
 the gravitationally bound system. It is found that the effect
 of cosmic expansion on the gravitationally bound systems on
 (or below) the galactic/cluster scales are insignificant. We
 refer to e.g.~\cite{bdsyspast,bdsysrecent,Nesseris:2004uj}
 for details. Thus, it is reasonable to adopt the adiabatic
 approximation and use the Minkowski instead of FRW metric as
 the background metric in the present work. In this case, the
 vierbein ${e^a}_\nu$ can be rewritten as a sum of flat
 component and small perturbed component, namely
\be{eq7}
{e^a}_\nu=\textup{diag}\,(1+\Phi,\,1-\Psi,\,1-\Psi,\,1-\Psi)\,,
\ee
where $\Phi$ and $\Psi$ are the gravitational potentials
($\Phi\ll 1$, $\Psi\ll 1$), and they are functions of spacetime
coordinates. From Eq.~(\ref{eq5a}), it is easy to get the
metric $g_{\mu\nu}$ as
\be{eq8}
g_{\mu\nu}=\textup{diag}\,(1+2\Phi,
\,-1+2\Psi,\,-1+2\Psi,\,-1+2\Psi)\,.
\ee
Note that $f(T)$ gravity is not a local Lorentz invariant
 theory~\cite{Li:2010cg} (we thank the referee for pointing
 out this issue). In general, the lack of local Lorentz
 symmetry implies that there is no freedom to fix any of the
 components of the tetrad~\cite{Li:2010cg}. The choice of
 tetrad is crucial, and different tetrads will lead to
 different field equations~\cite{Tamanini:2012hg}. The authors
 of~\cite{Tamanini:2012hg} suggested to speak of a ``good''
 tetrad if it imposes no restrictions on the form of $f(T)$.
 As mentioned above, we adopt the adiabatic approximation and
 use the Minkowski metric as the background metric. Note that
 for the Minkowski metric ${\rm diag}\,(+1,\,-1,\,-1,\,-1)$,
 the simplest choice for the background vierbein is given by
 ${\rm diag}\,(+1,\,+1,\,+1,\,+1)$, according to
 Eq.~(\ref{eq5a}). This motivates us to suggest the perturbed
 vierbein in Eq~(\ref{eq7}), which is actually the simplest
 background vierbein ${\rm diag}\,(+1,\,+1,\,+1,\,+1)$ plus
 small perturbed components $\Phi$, $\Psi$. In fact, the
 tetrad choice~(\ref{eq7}) is ``good'' according to the
 standard proposed in~\cite{Tamanini:2012hg}.
 On the other hand, the metric $g_{\mu\nu}$ in Eq~(\ref{eq8})
 obtained from the tetrad choice~(\ref{eq7}) via
 Eq.~(\ref{eq5a}) coincides with the well-known one considered
 in Newtonian gravity and the weak field limit of GR (see
 e.g.~\cite{mg1,fr1,Roshan2} and the references therein). This
 also supports the tetrad choice given in Eq.~(\ref{eq7}).
 In fact, the role played by the non-local Lorentz invariance
 in perturbation theory in the context of $f(T)$ gravity is
 far from clear, and the literature is quite confusing about
 this. However, in our particular case the Minkowski metric
 is perturbed at the first order, and the lack of local Lorentz
 invariance could not be a problem (we thank the referee for
 pointing out this issue). Anyway, let us move forward.

Substituting Eq.~(\ref{eq7}) into Eq.~(\ref{eq6}), we find
 the linearized vierbein field equations as
\bea
{E_0}^0 &=& -\partial_i \partial^i \Psi\, f_{T|0}
 = 4\pi G\rho \,, \label{eq9}\\[1mm]
{E_0}^i &=& \partial_0\partial^i\Psi\, f_{T|0}
 = 0 \,, \label{eq10}\\[1mm]
{E_i}^0 &=& \partial_j\partial^0\Psi\, \delta_i^j\, f_{T|0} = 0 \,,
 \label{eq11}\\[1mm]
{E_i}^j &=& \frac{1}{2}\partial_k
 \partial^j\left(\Psi-\Phi\right)\,\delta_i^k \,f_{T|0}
 = 4\pi G\partial_i \partial^j \pi^S
 \qquad (i\not=j) \label{eq12}\,, \\[1mm]
{E_i}^j &=& \left[-\delta_i^j \partial_0 \partial^0 \Psi +
 \frac{1}{2}\partial_k \partial^j \left(\Psi-\Phi\right)
 \delta_i^k-\frac{1}{2}\partial_k \partial^k \left(\Psi-\Phi\right)
 \delta_i^j\right] f_{T|0} = 4\pi G p \qquad (i=j)\label{eq13}\,,
\eea
where the indices $0$ and $i,\,j,\,k=1,\,2,\,3$ indicate
 the time component and space components of the spacetime
 coordinates, respectively; $\rho$ and $p$ are the density and
 pressure of the perfect fluid, respectively; $\pi^S$ is the scalar
 component of the anisotropic stress~\cite{Chen:2010va};
 $\partial_k$ denotes $\partial/\partial x^k$; the subscript
 ``$|0$'' means that the corresponding quantity is evaluated at
 $T=0$, for instance, $f_{T|0}=f_T\left(T=0\right)$. Note that
 the evaluation at $T=0$ in the dynamical equations is a consequence
 of the linearization process. We have adopted the adiabatic
 approximation, and hence the Minkowski spacetime is in the vacuum,
 i.e.~the zeroth order approximation of the energy-momentum
 tensor is vanishing ($\bar{\rho}=\bar{p}=0$). This makes the
 right-hand sides of Eqs.~(\ref{eq10}) and (\ref{eq11}) be
 zero, while $\rho$, $p$ are treated as the perturbations of
 the same order of $\Phi$, $\Psi$. On the other hand, one can
 find $f(T=0)=0$ by considering the zeroth order approximation
 of the vierbein field equations (namely the unperturbed equations).
 Actually, this is also a consequence of the adiabatic approximation
 (so the zeroth order approximation of the energy-momentum
 tensor is vanishing). In fact, the most popular $f(T)$ theories
 in the literature, e.g. the ones given by Eqs.~(\ref{eq16})
 and (\ref{eq17}), naturally satisfy $f(T=0)=0$. This also
 justifies the adiabatic approximation used in this work. From
 Eq.~(\ref{eq9}), if $\rho\not=0$, we see that $f_{T|0}\not=0$
 and the gravitational potential $\Psi$ must depend on the
 space coordinates (namely it cannot depend only on the time).
 Then, from Eqs.~(\ref{eq10}) and (\ref{eq11}), we find that
 $\partial_0\Psi=0$, and hence $\Psi$ must depend only on the
 space coordinates. If one adopts the zero-anisotropic-stress
 assumption, it is easy to see that $\Phi=\Psi$ and $p=0$ from
 Eqs.~(\ref{eq12}) and (\ref{eq13}). However, in the general
 cases we need not assume the vanishing of the anisotropic
 stress, and then $\Phi=\Psi$ and $p=0$ are not necessary. In
 the following discussions, we only use the Poisson's equation
 given in Eq.~(\ref{eq9}), without the assumptions $\Phi=\Psi$
 and $p=0$. The Poisson's equation of gravitational potential
 in $f(T)$ theory (namely Eq.~(\ref{eq9})) can be rewritten as
\be{eq18}
\Delta\Psi\equiv\nabla^2 \Psi= 4\pi G\alpha\rho\,,
\ee
where $\Delta\equiv\nabla^2$ is the well-known Laplacian, and
\be{eq18a}
\alpha\equiv 1/f_{T|0}\,.
\ee
As mentioned above, the results obtained in the present work do
 not rely on the specific forms of $f(T)$, and they are valid
 for all $f(T)$ theories satisfying $f(T=0)=0$
 and $f_T (T=0)\not=0$, if the adiabatic approximation and the
 weak field limit are considered. The information
 of the function $f(T)$ is mainly encoded in the parameter
 $\alpha$ given in Eq.~(\ref{eq18a}). If $f(T)=T$,
 or equivalently $\alpha=1$, our results reduce to the ones of
 GR. At first glance, one might redefine the gravitational
 constant as $G_{\rm eff}=\alpha G$, and then the Poisson's
 equation~(\ref{eq18}) becomes the one of GR. However, we note
 that this does not work if $\alpha\to 0$ or even $\alpha<0$,
 otherwise the effective gravitational constant becomes zero
 or negative. In fact, as is shown in Sec.~\ref{sec5}, if
 $\alpha\leq 0$, the disks are unconditionally stable. This is
 significantly different from the cases of
 $\alpha>0$ (especially $\alpha=1$ in GR), in which the disks
 can be stable only when some conditions are satisfied. Thus,
 we do not redefine the gravitational constant, so that the
 cases of $\alpha\leq 0$ can be included, which can lead to
 interesting results.


\section{Dispersion relation}\label{sec3}

In this section, we study the behavior of density waves in
 self-gravitating differentially rotating disks in $f(T)$
 theory by using density wave theory~\cite{Binney1,LS}. As
 mentioned in Sec.~\ref{sec1}, to the best of our knowledge,
 Lin and Shu~\cite{LS} are the pioneers who developed density
 wave theory for the first time. For convenience, in the
 present work, we mainly follow the modern formalism of
 density wave theory given in e.g.~\cite{Binney1}. We firstly
 calculate the gravitational potential by using the Poisson's
 equation~(\ref{eq18}), and then determine how the potential
 perturbation affects the equilibria of the self-gravitating
 system. We finally obtain the dispersion relations for
 gaseous and stellar systems in $f(T)$ theory, respectively.
 Note that the stability will be analyzed in the next section
 (Sec.~\ref{sec4}).

Since the gravitational force is long-range, galaxy can be
 regarded as a strongly coupled system. In principle, the wave
 patterns should be determined numerically. To be simple, in
 our analysis, we assume that the spiral galactic disks are
 tightly wound. In density wave theory, the long-range coupling
 can be neglected for tightly wound density waves~\cite{Binney1,LS}.
 Therefore, it is reasonable to adopt the tight-winding
 approximation (also known as Wentzel-Kramers-Brillouin~(WKB)
 approximation in the literature) to make the analytic
 derivations possible and simple. As mentioned in
 e.g.~\cite{Binney1,Roshan2}, the WKB approximation can be used
 in most cases without much loss of generality.

The stability analysis in self-gravitating fluid (gaseous) system is
 similar to the analysis in stellar system, because a fluid
 system is supported against gravity by gradients in the
 pressure $p$, while a stellar system is supported by gradients
 in the stress tensor~\cite{Binney1}. However, the stellar disk
 is more complicated than gaseous disk. It is easy to use the
 results of gaseous disk to analyze the difference between
 various theories. Since the response of a gaseous disk
 exhibits most of the important features that also occur in the
 stellar case, we find that it is helpful to consider spiral
 structure in gaseous disks before we tackle the
 more complicated stellar disks.


\subsection{Potential of a tightly wound spiral pattern}\label{sec3A}

Now, we consider an extremely thin disk whose surface density
 is $\Sigma$, occupying the plane $z=0$ and rotating in the $x$
 and $y$ directions (we use the Cartesian coordinate system
 $(x,\,y,\,z)$ here).~So, the Poisson's equation (\ref{eq18})
 can be rewritten as
\be{eq22}
\Delta\Psi= 4\pi G\alpha\Sigma\delta(z)\,.
\ee
The properties of an equilibrium system are described by the
 time-independent quantities. For the~Poisson's equation, we
 consider small perturbations of potential $\Psi$ and surface
 density $\Sigma$, which can be written as
 $\Psi=\Psi_0+\Psi_1$, $\Sigma=\Sigma_0+\Sigma_1$, where the
 subscript ``$0$'' denotes unperturbed (time-independent)
 components and the subscript ``$1$'' denotes perturbed
 components. Then, the linearized Poisson's equation~reads
\be{eq23}
\Delta\Psi_1= 4\pi G\alpha\Sigma_1\delta(z)\,.
\ee
To solve Eq.~(\ref{eq23}), it is convenient
 to adopt the following ansatz
 $\Sigma_1=\Sigma_a \exp{[i({\bf k \cdot x}-\omega t)]}$,
 $\Psi_1(z=0)=\Psi_a \exp{[i({\bf k \cdot x}-\omega t)]}$,
 where $\Sigma_a$ and $\Psi_a$ are constants, $\omega$ is the
 frequency and {\bf k} is the wave vector. Note that the
 perturbations should be real. If the complex functions
 $\Sigma_1$ and $\Psi_1(z=0)$ satisfy Eq.~(\ref{eq23}),
 $\Re(\Sigma_1)$ and $\Re(\Psi_1(z=0))$ also satisfy the
 equation, where $\Re(\xi)$ gives the real part of the complex
 function $\xi$. So, we allow $\Sigma_1$ and $\Psi_1(z=0)$ to
 be complex, with the understanding that the physical density
 is given by its real part instead. We refer
 to~\cite{Binney1} for detailed discussions.

Without loss of generality, we choose $x$-axis to be parallel
 to ${\bf k}$, so ${\bf k}= k{\bf e_x}$, where $k$ is the wavenumber
 and ${\bf e_x}$ is the unit vector parallel to $x$-axis. Since when
 $z\not=0$ we have $\Delta\Psi_1=0$, the potential perturbation can
 be rewritten as
\be{eq23a}
\Psi_1=\Psi_a\, e^{i({\bf k \cdot x}-\omega t) -|kz|} \,.
\ee
To relate $\Psi_a$ to $\Sigma_a$, we integrate Eq.~(\ref{eq23}) from
 $z=-\zeta$ to $z=+\zeta$, where $\zeta>0$ and $\zeta\to 0$.
 Since ${\partial^2\Psi_1}/{\partial x^2}$ and
 ${\partial^2\Psi_1}/{\partial y^2}$ are continuous at $z=0$,
 but ${\partial^2\Psi_1}/{\partial z^2}$ is not, we have
\be{eq24}
\lim _{\zeta\to 0}\int_{-\zeta}^{+\zeta}\frac{\partial^2\Psi_1}
 {\partial z^2}\,dz = \lim_{\zeta\to 0}\left.\frac{\partial
 \Psi_1}{\partial z}\right|_{-\zeta}^{+\zeta} = -2|k|\Psi_1 =
 \lim_{\zeta\to 0 }4\pi G\alpha\Sigma_1\int_{-\zeta}^{+\zeta}
 \delta(z)\,dz = 4\pi G \alpha\Sigma_1 \,,
\ee
where we have used Eq.~(\ref{eq23a}) in the second ``$=$''.
 So, it is easy to find that
\be{eq25}
\Psi_a = -\frac{2\pi G\alpha}{|k|}\,\Sigma_a\,.
\ee
If $\alpha\to 1$, Eq.~(\ref{eq25}) reduces to the Newtonian
 case as expected. It is worth noting that the derivation of
 Eq.~(\ref{eq25}) only involves the fundamental aspects of
 wave equations, regardless of whether the disks are uniformly
 rotating or differentially rotating. So, Eq.~(\ref{eq25}) is
 valid for both the uniformly and differentially rotating
 disks~\cite{Binney1}. Note also that Cartesian coordinates
 are used when we derive the gravitational potentials. One
 should be aware that we might switch to polar or cylindrical
 coordinates in the following derivations, although the
 transformation between different coordinate systems is easy.

The disks considered above are the uniformly rotating disks in
 which we can use general wave dynamics. If we instead consider
 the differentially rotating disks, the structure of spiral
 arms in the disks should be regarded as density
 waves, according to the density wave theory~\cite{Binney1}.

To be self-contained, here we briefly introduce the so-called
 tightly wound spiral arms~\cite{Binney1}. The spiral arm of
 galaxy can be viewed as a spiral curve which can be written as
 $\phi+g(R,t)=const.$, where $\phi$ is the azimuthal angle,
 $R$ is the distance from the center of galaxy to the point of
 the curve (we use the polar coordinate system here). We
 assume that the galaxy with $m$-fold symmetry has $m>0$ spiral
 arms ($m$ is a positive integer). So, the locations of all $m$
 arms are given by $m\phi+F(R,t)=const.$, where $F(R,t)=mg(R,t)$ is
 the shape function. The condition for tight winding is related
 to the so-called pitch angle $\vartheta$ of the arm which is given
 by $\cot \vartheta=\left |{kR}/{m}\right |$ at any radius $r$.
 When $\cot \vartheta =\left |{kR}/{m}\right | \gg 1$, namely
 the pitch angle $\vartheta$ is very small, we can say that the
 spiral arms are tightly wound.

The surface density of a zero-thickness  differentially
 rotating disk is the sum of an unperturbed component $\Sigma_0(R)$
 and a perturbed component $\Sigma_1(R, \phi, t)$. For a
 tightly wound spiral, according to density wave
 theory~\cite{Binney1}, $\Sigma_1(R, \phi, t)$ can be expressed
 as a form separating the rapid variations in density as one passes
 between arms from the slower variation in the strength of the
 spiral pattern as one moves along an arm. It is convenient to write
 the perturbed surface density $\Sigma_1$ as
\be{eq25a}
\Sigma_1= H(R,t)\,e^{i\,({m\phi} + F(R,\,t))} \,,
\ee
where $H(R,t)$ is a smooth function of radius $R$ that gives
 the amplitude of the spiral galaxy. Since the surface density
 oscillates rapidly around zero mean, the perturbed potential
 at a given location will be determined by the properties of
 the pattern within a few wavelengths of that location. We can
 replace the shape function $F(R,t)$ by its Taylor series in
 the vicinity of $(R_0, \phi_0)$,
 $F(R,t)\to F(R_0,t)+k(R_0,t)(R-R_0)$. The perturbed
 surface density $\Sigma_1$ can be rewritten as
\be{eq25b}
\Sigma_1\simeq\Sigma_a\,e^{ik(R_0,\,t)(R-R_0)}\,,
\ee
where
\be{eq25c}
\Sigma_a=H(R_0,t)\,e^{i({m\phi_0}+F(R_0,\,t))}\,.
\ee
From Eq.~(\ref{eq25b}), we find that the spiral wave closely
 resembles a plane wave with wave vector ${\bf k} =k {\bf e_R}$
 in the vicinity of $(R_0, \phi_0)$, where ${\bf e_R}$ is the
 unit vector parallel to $R$ direction. Note that
 the gravitational potential for uniformly rotating disks is
 given by Eq.~(\ref{eq25}). So, we can write the similar
 perturbed potential in differentially rotating disks as
\be{eq25d}
\Psi_1\simeq\Psi_a\, e^{ik(R_0,\,t)(R-R_0)}\,,
\ee
where $\Psi_a$ is given by Eq.~(\ref{eq25}). We refer to
 e.g.~\cite{Binney1} and references therein for a
 detailed introduction of density wave theory.

\vspace{-3mm} 


\subsection{Gaseous disks} \label{sec3b}

The stability analysis in self-gravitating fluid system is
 similar to the analysis in stellar system, because a fluid
 system is supported against gravity by gradients in the
 pressure $p$, while a stellar system is supported by gradients
 in the stress tensor~\cite{Binney1}. On the other hand, the
 stability analysis of fluid system has been studied previously
 in the literature. So, it is better to consider the response
 of tightly wound density waves of self-gravitating gaseous
 system at first, and then deal with the similar analysis
 of self-gravitating stellar system later.

For our aim, it is necessary to obtain the corresponding continuity
 equation and the Euler equation, which are two of the most
 important equations in fluid dynamics and are useful for the
 stability analysis of the gaseous system. We consider a
 zero-thickness disk occupying the plane $z=0$ (we use the
 cylindrical coordinate system $(R,\,\phi,\,z)$ here). In
 Newtonian gravity, the continuity equation and
 Euler equation are given by~\cite{Binney1}
\bea
\frac{\partial \rho}{\partial t}+\nabla\cdot (\rho\, {\bf v})
 &=& 0 \,,\label{eq19}\\[1.5mm]
\frac{\partial \bf{v}}{\partial t}+(\bf{v}\cdot\nabla)\,{\bf v}
 &=& -\frac{1}{\rho}\nabla p-\nabla\Psi \,,\label{eq20}
\eea
respectively, where the pressure $p$ and velocity $\bf v$ of
 the perfect fluid act only in the disk plane $z=0$, and the
 volume density $\rho$ can be replaced by the surface density
 $\Sigma$ for our aim. According to~\cite{Roshan2}, since only
 the potential $\Psi$ appears in the equation of motion of a
 test particle, the continuity equation and Euler equation in
 $f(T)$ theory take the same forms in Newtonian gravity or
 $f(R)$ theory, i.e.~Eqs.~(\ref{eq19}) and (\ref{eq20}),
 while the information of $f(T)$ theory is carried by the
 modified gravitational potential $\Psi$.

By using cylindrical coordinates, the continuity equation
 (\ref{eq19}) can be rewritten as
\be{eq26}
\frac{\partial\Sigma}{\partial t} +\frac{1}{R}\frac{\partial}
 {\partial R}\left(\Sigma \,R\,v_R\right) + \frac{1}{R}
 \frac{\partial}{\partial \phi}\left(\Sigma\,R\,v_\phi\right)=0 \,,
\ee
where $v_R$ and $v_\phi$ are the radial and azimuthal components of
 velocity, respectively. On the other hand, the radial and azimuthal
 components of Euler equation (\ref{eq20}) can be rewritten as
\bea
\frac{\partial v_R}{\partial t} + v_R\frac{\partial v_R}{\partial R}
 + \frac{v_\phi}{R}\frac{\partial v_R}{\partial \phi} -
 \frac{v_\phi^2}{R} &=& -\frac{\partial\Psi}{\partial R} - \frac{1}
 {\Sigma}\frac{\partial p}{\partial R}\,,\label{eq27}\\[1.5mm]
\frac{\partial v_\phi}{\partial t} + v_R\frac{\partial v_\phi}
 {\partial R} + \frac{v_\phi}{R}\frac{\partial v_\phi}{\partial\phi}
 + \frac{v_\phi\, v_R}{R} &=& -\frac{1}{R}\frac{\partial\Psi}
 {\partial R}-\frac{1}{\Sigma R}\frac{\partial p}{\partial\phi}
 \label{eq28} \,.
\eea
Since the gaseous disk is introduced to serve as a heuristic
 model of stellar system, we are free to choose a simple
 equation of state $p= K\Sigma^\gamma$, where $K$ and $\gamma$
 are constants. It is convenient to define a specific enthalpy
 $h\equiv\disp\int \Sigma^{-1} dp=\frac{\gamma}{\gamma-1}
 \,K\Sigma^{\gamma-1}$, and then Eqs.~(\ref{eq27}) and
 (\ref{eq28}) can be simplified to
\begin{align}
\frac{\partial v_R}{\partial t} + v_R\frac{\partial v_R}{\partial R}
 + \frac{v_\phi}{R}\frac{\partial v_R}{\partial \phi} -
 \frac{v_\phi^2}{R} &= -\frac{\partial}{\partial R} \left(\Psi
 +h\right) \,, \label{eq29}\\[1.5mm]
\frac{\partial v_\phi}{\partial t} + v_R\frac{\partial v_\phi}
 {\partial R} + \frac{v_\phi}{R}\frac{\partial v_\phi}{\partial\phi}
 + \frac{v_\phi\, v_R}{R} &= -\frac{1}{R}\frac{\partial}{\partial R}
 \left(\Psi+h\right)\label{eq30}\,.
\end{align}
As mentioned above, the equilibrium of system is determined by
 the time-independent quantities. For gaseous system, we can
 consider that the spiral wave is a small perturbation on
 axisymmetric disk. We can write surface density
 $\Sigma= \Sigma_0+\Sigma_1$, radial component of velocity
 $v_R= v_{R0}+v_{R1}= v_{R1}$, azimuthal component of velocity
 $v_\phi= v_{\phi 0}+v_{\phi 1}$, gravitational potential
 $\Psi= \Psi_0 +\Psi_1$ and enthalpy $h= h_0 +h_1$, where the
 subscript ``$0$'' denotes unperturbed quantities and the subscript
 ``$1$'' denotes small perturbed quantities. Note that in the
 equilibrium, we have $v_{R0}=0$ and
 $\partial\Psi_0/\partial \phi=\partial h_0/\partial \phi=0$.
 The linearized equations of Eqs.~(\ref{eq26}), (\ref{eq29})
 and (\ref{eq30}) are given by
\begin{align}
&\frac{\partial\Sigma_1}{\partial t}+\frac{1}{R}\frac{\partial}
 {\partial R}\left(\Sigma_0\, R \,v_{R1}\right) +
 \Omega\frac{\partial \Sigma_1}{\partial \phi} +
 \frac{\Sigma_0}{R}\frac{\partial v_{\phi 1}}{\partial \phi}
 = 0 \label{eq31}\,, \\[1.5mm]
&\frac{\partial v_{R1}}{\partial t}
 +\Omega\frac{\partial v_{R1}}{\partial \phi}-2\,\Omega\,v_{\phi 1}=
 -\frac{\partial}{\partial R}\left(\Psi_1 + h_1
 \right)\,,\label{eq32}\\[1.5mm]
&\frac{\partial v_{\phi 1}}{\partial t} + \Omega
 \frac{\partial v_{\phi 1}}{\partial \phi} +\left[\frac{d\left(
 \Omega R\right)}{dR}+\Omega\right] v_{R1} = -\frac{1}{R}
 \frac{\partial}{\partial \phi} \left(\Psi_{1} +h_{1}\right)
 \label{eq33}\,,
\end{align}
where $\Omega=v_{\phi 0}/R$ is the circular frequency. For
 differentially rotating disk, $\Omega$ is related to radial
 coordinate $R$. According to the similar
 analysis in Sec.~\ref{sec3A}, the ansatz for the solutions
 of Eqs.~(\ref{eq31}), (\ref{eq32}) and (\ref{eq33}) can be
 written as
\bea
\Psi_{1} &=& \Re\left[\Psi_{a}(R)\,e^{i(m \phi-\omega t)}
 \right] \label{eq36} \,, \\[0.5mm]
\Sigma_{1} &=& \Re\left[\Sigma_{a}(R)\,e^{i(m \phi-\omega t)}
 \right] \label{eq37} \,, \\[0.5mm]
v_{R1} &=& \Re\left[v_{Ra}(R)\,e^{i(m \phi - \omega t)}
 \right] \label{eq38} \,, \\[0.5mm]
v_{\phi 1} &=& \Re\left[v_{\phi a}(R)\,e^{i(m \phi - \omega t)}
 \right] \label{eq39} \,, \\[0.5mm]
h_{1} &=& \Re\left[h_a(R)\,e^{i(m \phi-\omega t)}\right]
 \label{eq40}\,,
\eea
where $m>0$. Substituting Eqs.~(\ref{eq36})---(\ref{eq40})
 into Eqs.~(\ref{eq32}) and (\ref{eq33}), we get
\begin{align}
v_{Ra} &= -\frac{i}{\Theta}\left[\left(m\Omega-\omega\right)
 \frac{d}{dR}\left(\Psi_a+h_a\right) + \frac{2m\Omega}{R}
 \left(\Psi_a+h_a\right)\right] \label{eq41}\,, \\[1mm]
v_{\phi a} &= \frac{1}{\Theta}\left[-2B\frac{d}{dR}\left(\Psi_a
 +h_a\right)+\frac{m\left(m\Omega-\omega\right)}{R}\left(\Psi_a+h_a
 \right)\right] \label{eq42}\,,
\end{align}
where
\begin{align}
B &= -\frac{1}{2} \left[\frac{d\left(\Omega R\right)}{dR}
 +\Omega\right] = -\Omega -\frac{R}{2} \frac{d\Omega}{dR} \,,
 \label{eq34} \\[1.5mm]
\Theta &=\kappa^2-\left(m\Omega-\omega\right)^2 \,,\label{eq43}
\end{align}
in which the epicyclic frequency $\kappa$ is defined by
\be{eq35}
\kappa^2 = R\,\frac{d\,\Omega^2}{dR}+4\Omega^{2}=-4B\Omega \,.
\ee
The linearized equation of state is given by
\be{eq44}
h_a= \gamma K\Sigma_a\Sigma_0^{\gamma-2}=v_s^2\Sigma_a/\Sigma_0 \,,
\ee
where the sound speed $v_s$ is given
 by $v_s^2 = \gamma K\Sigma_0^{\gamma-1}$.

In principle, the linearized continuity equation (\ref{eq31}),
 the expressions of velocity Eqs.~(\ref{eq41}), (\ref{eq42}),
 and the linearized equation of state Eq.~(\ref{eq44}) can be
 numerically solved to yield the global forms of the self-consistent
 density waves in a given disk~\cite{Bardeen1}. Instead, here we use
 the WKB approximation (namely $|kR|\gg m\geq 1$) mentioned
 above to get the analytic local solutions for density waves.
 The potential of a tightly wound wave can be written as
\be{eq45}
\Psi_a(R)=W(R)\,e^{iF(R)} = W(R)\,
\exp\left(i \int_0^R k\,dR\right)\,,
\ee
where $k= dF(R)/dR$ and $|kR|\gg1$. From Eqs.~(\ref{eq23}),
 (\ref{eq36}), (\ref{eq37}) and (\ref{eq44}), we see that
 $\Sigma_a$ and $h_a$ share the same factor $\exp{(iF(R))}$
 with $\Psi_a\,$. Therefore, we can write
 $d(\Psi_a+h_a)/dR = i k(\Psi_a+h_a)$ and then neglect the
 terms proportional to $(\Psi_a+h_a)/R$ in Eqs.~(\ref{eq41})
 and (\ref{eq42}) because $|kR|\gg 1$. So, Eqs.~(\ref{eq41})
 and (\ref{eq42}) become
\begin{align}
v_{Ra} &= \frac{k\left(m\Omega-\omega\right)\left(\Psi_a+h_a\right)}
 {\Theta} \label{eq46} \,,\\[1.5mm]
v_{\phi a} &= -\frac{2iBk\left(\Psi_a+h_a\right)}{\Theta}
 \label{eq47}\,.
\end{align}
For the similar reason, the continuity equation (\ref{eq31})
 can be rewritten as
\be{eq48}
\left(m\Omega-\omega\right)\Sigma_a+k\Sigma_0\,v_{Ra}=0\,.
\ee
Using Eqs.~(\ref{eq48}), (\ref{eq46}), (\ref{eq44}), (\ref{eq43}) and
 (\ref{eq25}), we find that the dispersion relation for
 a gaseous disk in the tight-winding limit is given by
\be{eq49}
(m\Omega-\omega)^2=k^2 v_s^2-2\pi G |k|\alpha\Sigma_0 + \kappa^2\,.
\ee
Note that our result is suited for differentially
 rotating disks. If $\kappa=2\Omega$, noting Eqs.~(\ref{eq35}),
 (\ref{eq34}) and $\Omega=v_{\phi 0}/R$, we see that Eq.~(\ref{eq49})
 reduces to the dispersion relation for uniformly rotating disks. On
 the other hand, if $\alpha \to 1$, namely $f_{T|0}\to 1$,
 this dispersion relation in $f(T)$ theory reduces to the one
 in Newtonian gravity~\cite{Binney1}. Since in the weak
 field limit GR reduces to Newtonian gravity, the dispersion
 relation in $f(T)$ theory also reduces to the one in GR when
 $\alpha \to 1$. This is not surprising, because
 $f(T)$ theory is equivalent to GR if $f(T)=T$, as is well
 known. It is worth noting that the corresponding dispersion
 relation in $f(R)$ theory~\cite{Roshan2} is quite different
 from the one in $f(T)$ theory, namely Eq.~(\ref{eq49}).


\subsection{Stellar disks} \label{sec3c}

In this subsection, we study the dispersion relation for
 stellar disks. As a fluid system, the dynamics of gaseous disk
 is determined by the continuity equation and Euler equation.
 However, the dynamics of stellar disk is instead described
 by the so-called collisionless Boltzmann equation. The
 collisionless Boltzmann equation in Newtonian gravity (the
 weak field limit of GR) reads
\be{eq21}
\frac{\partial D}{\partial t}+{\bf v}\cdot\nabla D-\nabla\Psi
 \cdot\frac{\partial D}{\partial {\bf v}} = 0 \,,
\ee
where {\bf v} is the stellar velocity and $D$ is
 the distribution function. As mentioned above, the stability
 analysis in stellar system is closely similar to the one in
 fluid system, because a fluid system is supported against
 gravity by gradients in the pressure $p\,$, while a stellar
 system is supported by gradients in the stress
 tensor~\cite{Binney1}. So, we could use the similar technique
 used in the analysis for gaseous disks to obtain
 the dispersion relation for stellar disks. We
 refer to~\cite{Binney1} for more detailed discussions.

The important step is to determine
 the perturbation $\bar{v}_{R1}$ induced by $\Psi_1$ in the
 mean radial velocity of the stars at a given
 point ($R,\,\phi$). If the unperturbed orbits are circular,
 namely the disk is quite ``cold'',  $\bar{v}_{R1}$ is given by
\be{eq50}
\bar{v}_{Ra} = \frac{m\Omega-\omega}{\Theta}\,k\Psi_a \,,
\ee
which could be obtained from Eq.~(\ref{eq46}) with $h_a=0$,
 since the disk would be dynamically identical to a gaseous
 disk with zero pressure~\cite{Binney1}. This formula is
 accurate unless the stellar epicycle amplitude is much
 smaller than the wavelength $2\pi/k$ of spiral pattern.
 Otherwise, Eq.~(\ref{eq50}) should be modified to
\be{eq51}
\bar{v}_{Ra}=\frac{m\Omega-\omega}{\Theta}\,k\Psi_a {\cal F}\,,
\ee
where ${\cal F} \leq 1$ is the reduction factor. For a given
 potential perturbation, the reduction factor describes how
 much the response to a spiral perturbation is reduced below
 the value for a cold disk. Then, we can calculate the response
 density $\Sigma_a$ once we have $\bar{v}_{Ra}$, since the
 so-called Jeans equation, ${\partial\nu}/{\partial t}+
 {\partial(\nu\,\bar{v}_i)}/{\partial x_i}=0$, is identical to
 the continuity equation of the gaseous disk. So, the analogue
 of Eq.~(\ref{eq48}) is given by
\be{eq51a}
(m\Omega-\omega)\Sigma_a + k\Sigma_0\, \bar{v}_{Ra} = 0 \,.
\ee
According to~\cite{Binney1}, the reduction factor $\cal F$
 in Newtonian gravity is given by
\be{eq53}
{\cal F} (s,\chi)=\frac{1-s^2}{\sin\left(\pi s\right)}\int_{0}^{\pi}
 e^{-\chi\left(1+\cos\tau\right)}\sin\left(s \tau\right)\,
 \sin\tau\,d\tau \,,
\ee
where
\bea
&&s=\frac{\omega-m\Omega}{\kappa} \,, \label{eq54}\\
&&\chi=\left(\frac{k\sigma_R}{\kappa}\right)^2 \label{eq55}\,,
\eea
in which $\sigma_R$ is the radial velocity dispersion. We refer
 to e.g.~\cite{Binney1} and references therein for a detailed
 introduction to the reduction factor ${\cal F}$. Note that
 ${\cal F}$ takes the same form in different gravity theories,
 because the gravitational potential does not appear in
 Eq.~(\ref{eq53}). Using Eqs.~(\ref{eq51}), (\ref{eq51a}),
 (\ref{eq53}), (\ref{eq43}) and (\ref{eq25}), we obtain
 the dispersion relation for stellar disks,
\be{eq52}
(m\Omega-\omega)^2=\kappa^2 -
 2\pi G |k|\alpha\Sigma_{0\,}{\cal F}(s,\chi)\,.
\ee
Again, if $\alpha \to 1$, namely $f_{T|0}\to 1$,
 this dispersion relation in $f(T)$ theory reduces to the one
 in Newtonian gravity (the weak field limit of GR)~\cite{Binney1}. It
 is worth noting that the corresponding dispersion relation in
 $f(R)$ theory~\cite{Roshan2} is quite different from the one
 in $f(T)$ theory, namely Eq.~(\ref{eq52}).

The dispersion relations Eqs.~(\ref{eq49}) and (\ref{eq52}) are
 the main equations for studying the density waves in disks.
 Although the WKB approximation is valid conditionally, it can
 be used in most cases without much loss of generality, as
 mentioned in e.g.~\cite{Binney1,Roshan2}. These dispersion
 relations could provide an invaluable guide for the numerical
 analysis of stability, and they establish the relation between
 wavenumber and frequency that is satisfied by a traveling wave
 as it propagates across the disk~\cite{Binney1}.

Comparing the dispersion relations in $f(R)$
 theory~\cite{Roshan2} and $f(T)$ theory for both gaseous and
 stellar disks, we find that the main difference between them
 is that the dispersion relations in $f(T)$ theory only relate
 to the first order derivative of $f(T)$, while the dispersion
 relations in $f(R)$ theory relate to not only the first but
 also the second order derivatives of $f(R)$. So, we suppose
 that it might be a possible hint to distinguish $f(T)$ theory
 from $f(R)$ theory.


 \begin{center}
 \begin{figure}[tb]
 \centering
 \vspace{-6.3mm} 
 \includegraphics[width=0.5\textwidth]{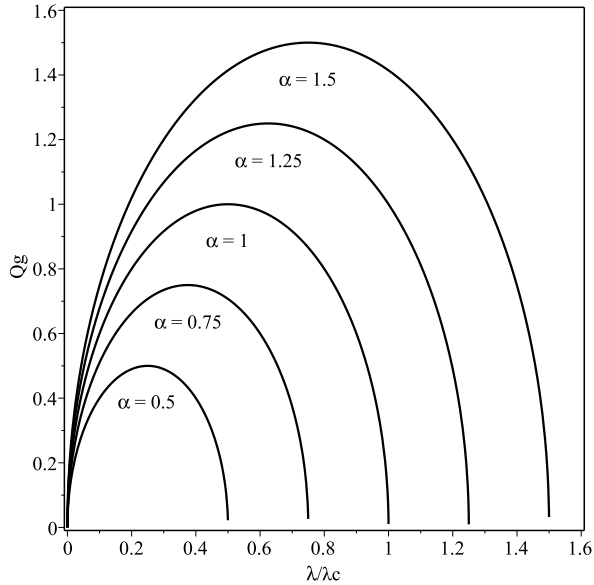}
 \caption{\label{fig1}
 The neutral stability curves for various $\alpha$ on the
 $Q_g-\lambda/\lambda_c$ plane according to Eq.~(\ref{eq57a})
 obtained in $f(T)$ theory for gaseous disk. Note that the
 curve with $\alpha=1$ corresponds to the one in GR and
 Newtonian gravity. See the text for details.}
 \end{figure}
 \end{center}


\vspace{-8.8mm} 


\section{Local stability}\label{sec4}

In the previous section, we obtained the dispersion relations
 Eqs.~(\ref{eq49}) and (\ref{eq52}) for gaseous and stellar
 disks in $f(T)$ theory, respectively. In this section, we
 apply these relations to determine whether a given disk is
 locally stable to axisymmetric perturbations. Note that the
 above analysis is mainly for tightly wound non-axisymmetric
 disturbances, namely $|kR/m|\gg 1$ and $m>0$. However, the
 results obtained above are still valid for the axisymmetric
 disturbances ($m=0$), so long as $|kR|\gg 1$~\cite{Binney1}.
 In the following, we try to obtain the local stability
 criteria in $f(T)$ theory for both gaseous and stellar
 disks.


\subsection{Gaseous disks} \label{sec4a}

At first, we consider the local stability of gaseous disks.
 For $m=0$, Eq.~(\ref{eq49}) becomes
\be{eq55a}
\omega^2=k^2 v_s^2 - 2\pi G |k|\alpha\Sigma_0 + \kappa^2\,.
\ee
The right-hand side of Eq.~(\ref{eq55a}) is real, so $\omega^2$
 should be real. If $\omega^2 > 0$, the frequency $\omega$ is
 real and hence the disk is stable
 (nb. Eqs.~(\ref{eq36})---(\ref{eq40})). On the other hand, if
 $\omega^2 < 0$, the disk is unstable. The neutral stability
 curve ($\omega^2=0$) is given by
\be{eq56}
k^2 v_s^2 - 2\pi G |k| \alpha\Sigma_0 + \kappa^2 = 0 \,,
\ee
which is a quadratic equation with respect to $|k|$ in fact.
 Noting that Eq.~(\ref{eq55a}) can be regarded as a parabola
 opening upward on the $\omega^2-|k|$ plane, if there is no
 real solution for the quadratic equation with respect to
 $|k|$ given in Eq.~(\ref{eq56}), we have $\omega^2>0$ for
 any $|k|$, and hence the gaseous disk is stable against all
 axisymmetric perturbations. Thus, by requiring that there is
 no real solution for Eq.~(\ref{eq56}), it is easy to obtain
 the condition for axisymmetric stability,
\be{eq57}
Q_g \equiv\frac{v_s\kappa}{\pi G\Sigma_0} > \alpha \,.
\ee
This is the local stability criterion in $f(T)$ theory for
 gaseous disk. In order to plot the curve of neutral stability
 given in Eq.~(\ref{eq56}), it is useful to introduce the
 longest unstable wavelength
 $\lambda_c = 4\pi^2 G \Sigma_0/\kappa^2$ and $y=\lambda/\lambda_c$,
 where $\lambda=2\pi/|k|$. So, we can recast Eq.~(\ref{eq56}) as
\be{eq57a}
Q_g= 2\sqrt{\alpha y - y^2} \,,
\ee
where $Q_g\equiv v_s\kappa/(\pi G\Sigma_0)$ is defined in
 Eq.~(\ref{eq57}). In Fig.~\ref{fig1}, we plot the neutral
 stability curves for various $\alpha$ on the
 $Q_g-\lambda/\lambda_c$ plane according to Eq.~(\ref{eq57a}).
 Noting that Eq.~(\ref{eq55a}) can be rewritten as
 \be{eq57b}
 Q_g^2-4\left(\alpha y-y^2\right)=
 \omega^2 \left(\frac{\kappa}{\pi G |k|\Sigma_0}\right)^2\,,
 \ee
 the disk is stable/unstable in the regions above/below the
 neutral stability curve for a given $\alpha$ on
 the $Q_g-\lambda/\lambda_c$ plane, respectively. Note that
 Eq.~(\ref{eq57}) reduces to the one in Newtonian gravity
 if $\alpha\to 1$.


\subsection{Stellar disks} \label{sec4b}

Let us turn to the local stability of stellar disks. In the case of
 axisymmetric perturbations ($m=0$), Eq.~(\ref{eq52}) becomes
\be{eq58}
\omega^2=\kappa^2 - 2\pi G |k| \alpha\Sigma_{0\,} {\cal F}
 \left(\omega/\kappa,\,\left({k\sigma_R}/{\kappa}\right)^2\right)\,.
\ee
Again, noting Eqs.~(\ref{eq36})---(\ref{eq40}), if $\omega^2>0$
 the disk is stable, and if $\omega^2<0$ the disk is unstable.
 The neutral stability curve ($\omega^2=0$) is given by
\be{eq58a}
\kappa^2 - 2\pi G |k| \alpha\Sigma_{0\,} {\cal F}
 \left(0,\,\left({k\sigma_R}/{\kappa}\right)^2\right) = 0 \,.
\ee
By using Eqs.~(\ref{eq53})---(\ref{eq55}) and the identity
 $\exp\left({z\cos\theta}\right) = \sum\limits_{n=-\infty}^{+\infty}
 I_n(z)\cos\left(n\theta\right)$~\cite{Binney1}, it can be recast as
\be{eq59}
\frac{|k|\sigma_R^2}{2\pi G\alpha\Sigma_0}=1-\exp\left(
 -\frac{k^2\sigma_R^2}{\kappa^2}\right)I_0\left(
 \frac{k^2\sigma_R^2}{\kappa^2}\right) \,,
\ee
where $I_0\left(\chi\right)$ is the modified Bessel function.
 If there is no real solution for Eqs.~(\ref{eq58a}) or (\ref{eq59})
 with respect to $|k|$, the stellar disk is stable against all
 axisymmetric perturbations. Note
 that Eqs.~(\ref{eq58})---(\ref{eq59}) are closely similar to
 the ones in Newtonian gravity, except that an additional
 factor $\alpha$ appears. Therefore, in analogy to the
 derivations in Newtonian gravity~\cite{Binney1}, one can obtain the
 modified local stability criterion in $f(T)$ theory as
\be{eq63}
Q_s \equiv \frac{\kappa\sigma_R}{3.36 G\Sigma_0} > \alpha \,.
\ee
Obviously, it reduces to the standard Toomre's criterion in
 Newtonian gravity if $\alpha\to 1$.

Note that the neutral stability curve determined
 by Eqs.~(\ref{eq58a}) or (\ref{eq59}) involves an integration
 function $\cal F$ given in Eq.~(\ref{eq53}) or a modified
 Bessel function $I_0$. So, it is not easy to solve the complicated
 equations given in Eqs.~(\ref{eq58a}) or (\ref{eq59}). Some
 complicated tricks are needed, as in e.g.~\cite{Roshan2}. However,
 the resulted curves are similar to the ones in Fig.~\ref{fig1} (see
 e.g.~\cite{Roshan2}), and hence we do not present them here.

It is of interest to briefly compare the stability criteria in
 $f(T)$ theory (Eqs.~(\ref{eq57}) and (\ref{eq63})) with the
 ones in $f(R)$ theory obtained in~\cite{Roshan2}. The main
 difference is that the stability criteria in $f(T)$ theory
 involve only the first order derivative of the function $f(T)$
 (encoded in the parameter $\alpha$), whereas the stability
 criteria in $f(R)$ theory involve not only the first but also
 the second order derivatives of the function $f(R)$ (encoded
 in the parameters $\alpha$ and $\beta$~\cite{Roshan2},
 respectively). This is mainly due to the fact that the
 equations of motion in $f(T)$ theory are 2nd order, whereas
 the ones of $f(R)$ theory are 4th order.


\section{Concluding remarks}\label{sec5}

In this work, we consider the stability of self-gravitating
 differentially rotating disks in $f(T)$ theory. At first, we
 get the Poisson's equation in the weak field limit, using the
 adiabatic approximation. Then, we obtain the gravitational
 potential in differentially rotating disks for a small
 perturbation. By studying the behavior of density wave, we
 obtain the dispersion relations for both gaseous and stellar
 disks. Finally, we study the local stability of both gaseous
 and stellar disks by applying the dispersion relations,
 and~obtain the modified Toomre's criteria (the
 local stability criteria) in $f(T)$ theory, i.e.
 Eqs.~(\ref{eq57}) and~(\ref{eq63}).

Comparing the local stability criteria in
 $f(R)$ theory~\cite{Roshan2} and $f(T)$ theory, we find that
 the main difference between them is that the local stability
 criteria in $f(T)$ theory only relate to the first order
 derivative of $f(T)$, while the local stability criteria
 in $f(R)$ theory relate to not only the first but also the
 second order derivatives of $f(R)$. So, we suppose that it
 might be a possible hint to distinguish $f(T)$ theory from
 $f(R)$ theory.

Let us observe the local stability criteria~(\ref{eq57}) and
 (\ref{eq63}) closely. If $\alpha\to 1$, namely $f_{T|0}\to 1$,
 both of Eqs.~(\ref{eq57}) and (\ref{eq63}) reduce to the
 standard Toomre's criteria in Newtonian gravity. This is not
 surprising, because $f(T)$ theory is equivalent to GR if $f(T)=T$,
 while GR reduces to Newtonian gravity in the weak field limit.
 However, if $\alpha\not=1$, Toomre's criterion should be modified.
 If $\alpha$ is larger/smaller than~$1$, a larger/smaller $Q_g$
 (compared with the one in Newtonian gravity or the weak field
 limit of GR) is needed to make the disk stable, respectively.
 This means that the disk needs larger/smaller pressure
 (compared with the one in Newtonian gravity or the weak
 field limit of GR) to resist the gravitational
 collapse~\cite{Binney1,Roshan2}. We consider that this might
 be potentially used to distinguish $f(T)$ theory from GR
 observationally.

It is interesting to consider the case of $\alpha\to 0$, namely
 $f_{T|0}\to\infty$. For instance, the typical form of
 $f(T)$ in Eq.~(\ref{eq17}) which has been extensively
 considered in the literature, can satisfy not only $f(T=0)=0$
 (required by the adiabatic approximation as mentioned in
 Sec.~\ref{sec2B}), but also $f_{T|0}=f_T(T=0)\to\infty$. In
 the case of $\alpha\to 0$ (namely $f_{T|0}\to\infty$),
 noting Eqs.~(\ref{eq55a}) and (\ref{eq58}), it is easy to see
 that $\omega^2>0$ holds unconditionally, and hence the disks
 are always stable. This can also be found by considering the local
 stability criteria~(\ref{eq57}) and (\ref{eq63}), which are
 satisfied unconditionally if $\alpha\to 0$. Therefore, the
 disks are unconditionally stable in e.g. the $f(T)$ theory
 given in Eq.~(\ref{eq17}). Similarly, the same arguments also
 hold in the case of $\alpha<0$, namely $f_{T|0}<0$. For
 instance, $f(T)=T+c_1(1-\exp\left(c_2T\right))$ can satisfy
 $f(T=0)=0$ (required by the adiabatic approximation), and
 $f_{T|0}<0$ if the model parameters satisfy $c_1 c_2>1$. The disks
 are also unconditionally stable. So, the kinds of $f(T)$
 theories with $\alpha\leq 0$ might be observationally
 tested on (or below) the galactic scales.

Note that in our analysis, we restrict ourselves to the
 adiabatic approximation, and hence the Minkowski metric can be
 taken as the background metric. It means that the evolution of
 the universe is very slow in comparison with local dynamics,
 and the physics of gravitationally bound systems (such as
 galaxies, clusters, or planetary systems) which are small
 compared to the radius of curvature of the cosmological
 background is essentially unaffected by the expansion of the
 universe. The adiabatic approximation has been supported by
 many quantitative studies (e.g.~\cite{bdsyspast,bdsysrecent,
 Nesseris:2004uj}) since 1945 at least. However, it has been
 challenged recently, and the debate is not completely settled
 by now. On the other hand, as mentioned in Sec.~\ref{sec2B},
 $f(T=0)=0$ is required by the adiabatic approximation. However, not
 all $f(T)$ theories satisfy this requirement in fact. Based on
 the above two arguments, it is also plausible to give up
 the adiabatic approximation, and choose e.g. the FRW metric as the
 background metric. In this case, the requirement $f(T=0)=0$
 is not necessary. However, as mentioned in e.g.~\cite{Chen:2010va},
 another constraint $f_{TT|0}=f_{TT}(T=0)=0$ can be imposed by
 the requirement of no anisotropic stress (of course, this
 constraint can also be given up, by allowing anisotropic
 stress). On the other hand, since the background metric is
 chosen to be e.g. the FRW metric or the interpolating
 metric~\cite{Nesseris:2004uj,Wei:2012ct} in this case, all the
 derivations in this work will become fairly complicated, and
 hence the modified local stability criteria should also
 become complicated accordingly. We also stress that the tetrad
 selection and the perturbative treatment could become
 significantly cumbersome. We leave this issue as an open question.

Another important assumption used in this work is the WKB
 approximation (namely the tight-winding approximation). As
 mentioned above, a natural criterion for the validity of
 the WKB approximation for axisymmetric waves is $|kR|\gg1$,
 which actually corresponds to $\lambda/R\ll 2\pi$. In fact,
 most of the numerical experiments require
 $\lambda/R\lesssim 2$ for the accuracy of the WKB results.
 So, the WKB approximation is valid in the solar neighborhood
 at least, and can be used in many galactic disks. However,
 the WKB approximation cannot be applied to the loose wound
 spiral structures. In this case, there are no analytic
 methods to determine the stability of a given disk to small
 perturbations, and the numerical methods should be used
 instead.


\section*{ACKNOWLEDGEMENTS}
We thank the anonymous referees for quite useful comments and
 suggestions, which helped us to improve this work. S.L.L. is
 grateful to Profs. Shuang-Qing~Wu, Rafael~Ferraro and
 Hong~L\"{u} for useful discussions and communications. We
 thank Zu-Cheng~Chen, Ya-Nan~Zhou, Xiao-Bo~Zou, Hong-Yu~Li and
 Dong-Ze~Xue for kind help and discussions. This work was supported
 in part by NSFC under Grants No.~11575022 and No.~11175016.

\renewcommand{\baselinestretch}{1.0}


\end{document}